\documentclass[fleqn,usenatbib]{mnras}

\usepackage{newtxtext,newtxmath}

\usepackage[T1]{fontenc}

\DeclareRobustCommand{\VAN}[3]{#2}
\let\VANthebibliography\thebibliography
\def\thebibliography{\DeclareRobustCommand{\VAN}[3]{##3}\VANthebibliography}

\usepackage{graphicx}	
\usepackage{amsmath}	

\title[Transit Spectroscopy of K2-33b]{Transit Spectroscopy of K2-33b with Subaru/IRD: Spin-Orbit Alignment and Tentative Atmospheric Helium}

\author[T. Hirano et al.]{%
Teruyuki Hirano,$^{1,2,3}$\thanks{E-mail: hd17156b@gmail.com (TH)}
Eric Gaidos,$^{4}$
Hiroki Harakawa,$^{5}$
Klaus W. Hodapp,$^{6}$
Takayuki Kotani,$^{1,2,3}$
\newauthor
Tomoyuki Kudo,$^{5}$
Takashi Kurokawa,$^{1,7}$
Masayuki Kuzuhara,$^{1,2}$
Andrew W. Mann,$^{8}$
Jun Nishikawa,$^{2,3,1}$
\newauthor
Masashi Omiya,$^{1,2}$
Takuma Serizawa,$^{7,2}$
Motohide Tamura,$^{9,1,2}$
Pa Chia Thao,$^{8}$
Akitoshi Ueda,$^{2}$
\newauthor
Sebastien Vievard$^{5,1}$
\\
$^{1}$Astrobiology Center, 2-21-1 Osawa, Mitaka, Tokyo 181-8588, Japan\\
$^{2}$National Astronomical Observatory of Japan, 2-21-1 Osawa, Mitaka, Tokyo 181-8588, Japan\\
$^{3}$Department of Astronomical Science, School of Physical Sciences, The Graduate University for Advanced Studies (SOKENDAI), 2-21-1, Osawa, Mitaka, Tokyo, 181-8588, Japan\\
$^{4}$Department of Earth Sciences, University of Hawai'i at M\"{a}noa, Honolulu, HI 96822, USA\\
$^{5}$Subaru Telescope, 650 N. Aohoku Place, Hilo, HI 96720, USA\\
$^{6}$University of Hawaii, Institute for Astronomy, 640 N. Aohoku Place, Hilo, HI 96720, USA\\
$^{7}$Institute of Engineering, Tokyo University of Agriculture and Technology, 2-24-16, Nakacho, Koganei, Tokyo, 184-8588, Japan\\
$^{8}$Department of Physics and Astronomy, The University of North Carolina at Chapel Hill, Chapel Hill, NC 27599, USA\\
$^{9}$Department of Astronomy, Graduate School of Science, The University of Tokyo, 7-3-1 Hongo, Bunkyo-ku, Tokyo 113-0033, Japan
}

\date{Accepted XXX. Received YYY; in original form ZZZ}

\pubyear{2024}

\begin{document}
\label{firstpage}
\pagerange{\pageref{firstpage}--\pageref{lastpage}}
\maketitle

\begin{abstract}
Exoplanets in their infancy are ideal targets to probe the formation and evolution history of planetary systems, including the planet migration and atmospheric evolution and dissipation. In this paper, we present spectroscopic observations and analyses of two planetary transits of K2-33b, which is known to be one of the youngest transiting planets (age $\approx 8-11$ Myr) around a pre-main-sequence M-type star. Analysing K2-33's near-infrared spectra obtained by the IRD instrument on Subaru, we investigate the spin-orbit angle and transit-induced excess absorption for K2-33b. We attempt both classical modelling of the Rossiter-McLaughlin (RM) effect and Doppler-shadow analyses for the measurements of the projected stellar obliquity, finding a low angle of $\lambda=-6_{-58}^{+61}$ deg (for RM analysis) and $\lambda=-10_{-24}^{+22}$ deg (for Doppler-shadow analysis). In the modelling of the RM effect, we allow the planet-to-star radius ratio to float freely to take into account the possible smaller radius in the near infrared, but the constraint we obtain ($R_p/R_s=0.037_{-0.017}^{+0.013}$) is inconclusive due to the low radial-velocity precision.  Comparison spectra of K2-33 of the 1083 nm triplet of metastable ortho-He I obtained in and out of the 2021 transit reveal excess absorption that could be due to an escaping He-rich atmosphere.  Under certain conditions on planet mass and stellar XUV emission, the implied escape rate is sufficient to remove an Earth-mass H/He in $\sim$1 Gyr, transforming this object from a Neptune to a super-Earth. 
\end{abstract}




\section{Introduction}


Planets in their infancy are invaluable to the study of planet formation and evolution. In particular, Neptune- to Jupiter-size planets younger than $\sim 100$ million years (Myr) can provide important clues to their evolution, as pronounced changes to their orbits (e.g., scattering) and atmospheres have been predicted to take place on this timescale \citep{Winn2015,Owen2019}.  Since the {\it K2} mission \citep{2014PASP..126..398H}, dozens of young transiting exoplanets have been identified, some of which are amenable to orbital and atmospheric characterization by transit spectroscopy \citep[e.g.,][]{2019ApJ...885L..12D, 2019ApJ...880L..17N, 2020Natur.582..497P,2023AJ....165...23T}.

Specifically, measurement of projected stellar obliquity with respect to the orbit plane (denoted by $\lambda$) tests scenarios for the evolution of planets onto close-in orbits, e.g. migration within a primordial disk (producing a low obliquity) vs. planet-planet scattering or a secular resonance such as Kozai-Lidov followed by tidal circularization (producing a high obliquity). Measurement of obliquities in young systems is especially valuable since the realignment of the stellar spin and planetary orbit is expected to occur on a Gyr timescale \citep[e.g.,][]{2010ApJ...718L.145W,2012ApJ...757...18A}, after which the different scenarios are indistinguishable. Moreover, while migration within a disk must obviously have been completed while the disk was extant ($\lesssim$10 Myr), both planet-planet scattering and the Kozai-Lidov mechanism can operate after the disk is gone, on timescales of $\gtrsim 1-10$ Myr \citep[e.g.,][]{2007ApJ...669.1298F,2011ApJ...742...72N,2016MNRAS.458.4345M}.

The technique of transmission spectroscopy can probe the atmospheres of transiting planets if the atmospheric scale height is sufficiently large, e.g. by a high planet equilibrium temperatures and/or low molecular weight atmosphere, and especially if the atmospheres are escaping in a wind.  Exceptionally young, close-in planets that are heavily irradiated by their host stars may have extended, escaping H/He-rich atmospheres.  Atmospheric escape driven by stellar high-energy irradiation over $\sim 100$ Myr is thought to be responsible for the formation of the gap between the radius distributions of super-Earths and sub-Neptunes 
\citep{2017ApJ...847...29O,2017AJ....154..109F} as well as the ``Neptune" desert close to stars, although there are other mechanisms proposed to explain those observed properties
of exoplanets including the ``core-powered mass-loss" \citep[e.g.,][]{2018MNRAS.476..759G, 2022MNRAS.516.4585G}.  
While the allowed transitions of H~I (e.g., Lyman and Balmer series) are either at vacuum wavelengths not accessible from the ground or are not excited at typical wind temperature, the strong 1083.3 nm triplet of metastable He~I from  its ortho- electronic state has been used to detect or constrain the extended atmospheres or winds of several young gas-rich exoplanets \citep{ Vissapragada2021,2022MNRAS.509.2969G,Zhang2022a,Zhang2022b,Zhang2023,Gaidos2023}.  


K2-33b,  super-Neptune-sized planet in a $5.42-$day orbit, is the youngest transiting exoplanet known to date \citep{2016AJ....152...61M,2016Natur.534..658D}. 
Its host is a pre-main-sequence low-mass star, residing in the 
8-11~Myr Upper-Scorpius star forming region \citep{2016AJ....152...61M}.
Despite the star's youth, there is no evidence for a circumstellar disk.   One aspect of interest is the planet's apparent radius: the planet size was measured as $R_p=5.04_{-0.37}^{+0.34}\,R_\oplus$ based on transit observations with {\it K2} ($\approx0.6$ $\mu$m) and the ground-based MEarth telescope array (0.84$\mu$m) \citep{2016AJ....152...61M}, but a recent study using transit observations at 1.1-1.7 $\mu$m with the Wide Field Planetary Camera (WFPC3) on the {\it Hubble} Space Telescope and 3.6 and 4.5 $\mu$m with the IRAC camera on the {\it Spitzer} telescope indicate a 30\% smaller radius than derived from transit observations optical wavelengths \citep{2023AJ....165...23T}, possibly due to photochemical haze in an extended atmosphere, or star spots.  \citet{2022ApJ...940L..30O} proposed an alternative explanation involving wavelength-dependent scattering in a circumplanetary dust ring, a scenario which can better explain the transmission  spectrum of the planet.   Further spectroscopic characterisation of K2-33b would help us solve the puzzle
of the radius discrepancy, as well as better understand the origin and evolution of this infant planet.  

Here, we present high-resolution near-infrared spectroscopy of transits of K2-33b using the InfraRed Doppler (IRD) instrument \citep{2012SPIE.8446E..1TT, 2018SPIE10702E..11K}  on the Subaru telescope, with the aim of  investigating the projected stellar obliquity and the extended helium atmosphere of the planet.  The large radius of K2-33b for its insolation level \citep{2016AJ....152...61M,2018AJ....155..127H} makes the planet a particularly appropriate target for such a study, and IRD has also been successfully used to search for extended atmospheres around other young planets  \citep{2020MNRAS.495..650G, 2020MNRAS.498L.119G, 2020ApJ...899L..13H, 2022MNRAS.509.2969G}.  

\section{IRD Spectroscopy for K2-33} \label{sec:observation}

We obtained spectra of K2-33 during transits of ``b" using IRD, a fiber-fed spectrograph developed for precise RV measurements in the near infrared ($970-1730$ nm) with the spectral resolution of $R\approx 70,000$ \citep{2018SPIE10702E..11K}. During the science exposures, we also injected light from the laser-frequency comb (LFC) into IRD as a simultaneous wavelength reference. Typical integration time was 900--1500 sec for each frame. 
We obtained 17 and 19 spectra on UT 2020 June 8 and UT 2021 April 4, respectively, covering the full transit but only limited out-of-transit baseline on both nights. Since the first two frames on UT 2020 June 8 were severely affected by poor weather (i.e., counts less than 100 e$^-$ at 1000 nm), we discarded them from the subsequent analyses.

 We reduced the raw IRD frames with IRAF \citep{1993ASPC...52..173T} and our custom code,  and extracted wavelength-calibrated, one-dimensional spectra for the star
 and LFC separately. K2-33's reduced spectra typically had a signal-to-noise (S/N) ratio 
of 30--45 per pixel at $1000$ nm. We analysed those reduced spectra and 
computed relative RVs for individual frames using IRD's standard pipeline for precise RV measurements;
we generated a telluric-free template spectrum for K2-33 using multiple observed
spectra, to which the relative RVs were measured by the forward modelling technique 
\citep{2020PASJ...72...93H}. 
Instrumental drifts were corrected by taking into account the variations in the
instantaneous instrumental profile of the spectrograph. 
For each frame, RVs were calculated for spectral segments each covering $\approx 2$ nm, 
and we derived the final RV for each frame
by taking the weighted mean of the segment RVs. 
Those relative RVs and their errors are summarized in Table~\ref{tab:ird}. 

\begin{table}
\centering
\caption{Relative RVs of K2-33}\label{tab:ird}
\begin{tabular}{rcc}
\hline\hline
BJD (TDB) & Value (m s$^{-1}$) & Uncertainty (m s$^{-1}$)   \\\hline
2459008.860805 & $21.02$ & 55.20 \\ 
2459008.884388 & $19.81$ & 13.35 \\ 
2459008.898993 & $10.88$ & 13.49 \\ 
2459008.913439 & $22.62$ & 12.43 \\ 
2459008.927990 & $-2.04$ & 14.11 \\ 
2459008.942657 & $5.82$ & 19.02 \\ 
2459008.957160 & $-1.42$ & 12.32 \\ 
2459008.971873 & $-12.16$ & 10.97 \\ 
2459008.987250 & $14.36$ & 11.78 \\ 
2459008.999268 & $-13.20$ & 11.63 \\ 
2459009.011491 & $-4.30$ & 12.85 \\ 
2459009.023715 & $3.32$ & 12.69 \\ 
2459009.035956 & $-14.37$ & 13.33 \\ 
2459009.047963 & $-32.79$ & 14.77 \\ 
2459009.057836 & $-36.36$ & 22.30 \\ 
2459328.921323 & $-35.42$ & 13.61 \\ 
2459328.932023 & $-34.19$ & 14.02 \\ 
2459328.943469 & $-38.27$ & 13.48 \\ 
2459328.954914 & $-20.53$ & 13.84 \\ 
2459328.965615 & $-28.75$ & 14.16 \\ 
2459328.976318 & $12.08$ & 14.32 \\ 
2459328.987013 & $-26.62$ & 12.53 \\ 
2459328.997733 & $-18.21$ & 11.88 \\ 
2459329.008434 & $-28.36$ & 11.53 \\ 
2459329.019128 & $-40.08$ & 13.70 \\ 
2459329.030025 & $-51.73$ & 16.27 \\ 
2459329.040720 & $28.15$ & 25.93 \\ 
2459329.051420 & $-23.49$ & 18.32 \\ 
2459329.062114 & $-45.70$ & 13.39 \\ 
2459329.072851 & $-36.24$ & 12.45 \\ 
2459329.083542 & $-65.14$ & 13.30 \\ 
2459329.094243 & $-55.36$ & 12.83 \\ 
2459329.116389 & $-38.48$ & 15.00 \\ 
2459329.127092 & $-24.61$ & 15.89 \\ 
2459329.137802 & $-26.12$ & 15.28 \\ 
\hline
\end{tabular}
\end{table}


\section{Measurements of the Stellar Obliquity} \label{sec:obliquity}

In this section, we present analyses of IRD data to measure the stellar obliquity. 
To model the spectroscopic transits of K2-33b, we first derived the central transit times ($T_c$) of our transit observations by IRD. 
Since an updated ephemeris was not given in the 
follow-up paper by \citet{2023AJ....165...23T}, we combined the ephemeris in the discovery paper
\citep{2016AJ....152...61M} with the transit times by Spitzer observations ($T_c=2457701.57374\pm 0.00091$; 
A. Mann, private communications) and obtained an updated period of 
$P=5.424871\pm0.000010$ days for K2-33b. Using this revised period, we estimated the 
predicted transit times of IRD observations as $T_c=2459008.9676 \pm 0.0026$ and 
$2459329.0350 \pm 0.0032$ for the 2020 and 2021 transits, respectively. 
When deriving the system parameters below, we take into account this uncertainty
in $T_c$. 
We note that no evidence of TTVs for K2-33b was reported in any of the past photometric
data sets (A. Mann, private communications).


\subsection{Modelling of the RM Effect} \label{sec:RM}
The RM effect has been routinely used to measure the projected stellar obliquity 
$\lambda$, which is manifested as an anomalous RV variation during transits 
\citep[e.g.,][]{2000A&A...359L..13Q,2005ApJ...631.1215W,2009ApJ...690....1O}. 
K2-33b is a challenging target for this type of observations due to its shallow transit
and faintness of the host star ($V\approx15.7$, $J=11.1$). Besides, K2-33b is reported to exhibit a chromatic
variation of the transit depth (or $R_p/R_s$), possibly originating from a geometrically thick, hazy atmosphere \citep{2023AJ....165...23T} or planetary ring \citep{2022ApJ...940L..30O}. 
The uncertainty in $R_p/R_s$ leads to
a systematic error in $\lambda$ when it is combined with the uncertainty in $v\sin i$
as well as the impact parameter $b$.

\begin{figure}
\centering
\includegraphics[width=8.5cm]{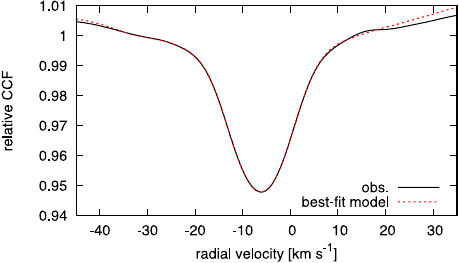}
\caption{Normalized CCFs between K2-33's IRD spectrum and the template spectrum of GJ 436. 
The black solid line represents the observed CCF while the red dashed line corresponds
to the best-fit theoretical model with $v\sin i=8.62$ km s$^{-1}$. 
}
\label{fig:vsini}
\end{figure}
In order to break the degeneracy between $v\sin i$ and $R_p/R_s$ and better
model the RV anomaly due to the RM effect, we first estimated $v\sin i$
from the IRD spectrum. Following the procedure in \citet{2020ApJ...899L..13H}, we computed
the mean cross-correlation function (CCF) between an observed IRD spectrum
of K2-33 and a template spectrum of a similar spectral type (i.e., GJ 436, an M3 star). 
CCF was computed for each spectral order of an IRD spectrum, and 
they were averaged to obtain the mean CCF. The black solid line in 
Figure \ref{fig:vsini}
indicates the resulting CCF for K2-33 (on UT 2021 April 24). 
To compare it with model CCFs, we generated a number of mock IRD
spectra with differing $v\sin i$ by convolving the observed IRD spectrum of 
GJ 436 \citep[$v\sin i<0.5$ km,][]{2018Natur.553..477B} with the rotation plus macroturbulence broadening kernel \citep{2011ApJ...742...69H}, in which we assumed
the macroburbulent velocity of $\zeta=1$ km s$^{-1}$. 
We then computed the CCF for each mock spectrum as for the observed
spectrum of K2-33. Using this set of mock CCFs with various $v\sin i$ values,
the observed CCF was fitted by interpolation to estimate $v\sin i$. 
We obtained $v\sin i = 8.62 \pm 0.04$ km s$^{-1}$, and the best-fit CCF model
is overplotted in Figure \ref{fig:vsini} by the red dashed line. 
The uncertainty in $v\sin i$ (0.04 km s$^{-1}$) is small, which only accounts for the 
statistical error in the fit and is likely underestimated principally due to uncertainty in the macroturbulence velocity. 
In what follows, we introduce a systematic error of $0.4$ km s$^{-1}$ for $v\sin i$ based on 
the difference between our $v\sin i$ value and that derived in \citet{2016AJ....152...61M}.

\begin{figure}
\centering
\includegraphics[width=8.5cm]{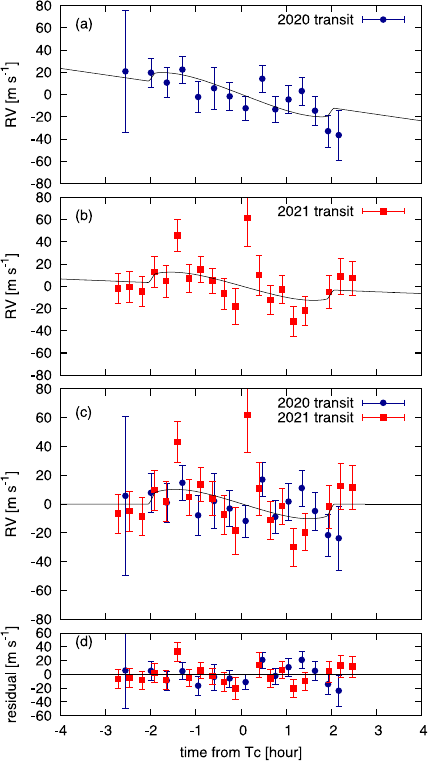}
\caption{RV variations by Subaru/IRD for the 2020 (panel (a)) and 2021 (panel (b)) transits. 
Panel (c) plots the two transit observations after subtracting the RV baseline for each dataset. 
In panels (a)--(c), the best-fit theoretical models are overplotted by the black solid lines. 
The RV residuals around the best-fit model are shown in panel (d). 
}
\label{fig:RM}
\end{figure}
We analysed the observed RVs from the two transit nights. 
Since the overall slopes of the RV baseline differ significantly between the two 
transit nights (Figure \ref{fig:RM}), we adopted two different baseline
parameters, introduced as an instantaneous Keplerian RV semi-amplitude
$K$ and RV offset parameter $\gamma$ for each night, which empirically 
represent the RV baselines on the transit nights. 

\begin{table*}
\centering
\caption{Result of the RM Analysis}\label{hyo6}
\begin{tabular}{lccc}
\hline\hline
Parameter & Value (RV modelling) & Value (Doppler shadow) & Prior  \\\hline
$v\sin i$ (km s$^{-1}$) & $8.60\pm0.40$ & $8.61_{-0.36}^{+0.30}$ & $\mathcal{G}$(8.6, 0.4) \\
$\lambda$ (degree) & $-6_{-58}^{+61}$ & $-10_{-24}^{+22}$ & $\mathcal{U}$ \\
$T_{c}$ ($\mathrm{BJD}-2454833$) & $4496.0351 \pm 0.0031$ & $4496.0349_{-0.0033}^{+0.0031}$ & $\mathcal{G}$(4496.0350, 0.0032) \\
$a/R_s$ & $10.31_{-0.29}^{+0.25}$ & $10.31_{-0.31}^{+0.23}$ & $\mathcal{G}$(10.29, 0.39) \\
$b$ & $0.17_{-0.12}^{+0.14}$ & $0.20_{-0.13}^{+0.14}$ & $\mathcal{U}$ \\
$R_p/R_s$ & $0.037_{-0.017}^{+0.013}$ & 0.0367 (fixed) & $\mathcal{U}$ \\
\hline
\end{tabular}
\end{table*}

We employed the RM velocity anomaly model by \citet{2011ApJ...742...69H}, which was derived in 
cases that observed RVs are extracted by the forward-modelling 
(template-matching) technique as for IRD's RVs. To take into account
the chromatic dependence of the transit depth \citep{2023AJ....165...23T}, we allowed
$R_p/R_s$ to float freely, although a single $R_p/R_s$ value was
fitted in the whole observing band covered by IRD ($Y$, $J$, and $H$). The fitting parameters
in our RM model are $\lambda$, $v\sin i$, $T_c$, the scaled semi-major
axis $a/R_s$, the transit impact parameter $b$, and $R_p/R_s$ in addition
to the baseline parameters for each transit data set described above. 
The limb-darkening parameters $u_1$ and $u_2$ for the quadratic law
are required in the RM model. Considering the low precision of the individual 
RV data points, we fixed those parameters at $u_1=0.09$ and $u_2=0.35$
based on the theoretical values for a $T_\mathrm{eff}=3500\,$K, $\log g=4.0$ star by \citet{2013A&A...552A..16C} for the $J$-band. 
To ensure the convergence of the fit, we imposed Gaussian priors of
$v\sin i=8.6\pm0.4$ km s$^{-1}$, $T_c-2454833=4496.0350\pm0.0032$, $a/R_s=10.29\pm0.39$, 
and the full transit duration of $T_{14}=4.08\pm0.07$ hours, based on the above 
discussion and literature values \citep{2016AJ....152...61M}. 
The last prior on $T_{14}$ was introduced in place of a prior on 
$b$, which was reported to have highly asymmetric errors (i.e., $b=0.16_{-0.11}^{+0.19}$).

Implementing the MCMC analysis \citep{2016ApJ...825...53H} in which the two transit data sets
are simultaneously modeled and fitted, we derived the system parameters
for K2-33b. The result of this analysis is presented in Table \ref{hyo6}. 
The stellar obliquity $\lambda$ was found to be $-6_{-58}^{+61}$ degrees
(the $68\,\%$ confidence interval), 
which is compatible with spin-orbit alignment, but alignment/misalignment is 
inconclusive from this RM analysis due to the large uncertainty. 
The low $R_p/R_s$ value of $0.037_{-0.017}^{+0.013}$ is more consistent 
with the HST result ($R_p/R_s\approx 0.0367$ for the HST white light curve at $1.088-1.68\,\mu$m)
than those reported by optical transit observations \citep[$R_p/R_s\approx 0.04735$ 
for K2 and $0.0489$ for MEarth,][]{2023AJ....165...23T}, although its large uncertainty makes
it inconclusive, too.  

To check for the dependency of the derived parameters ($\lambda$ in particular) on the choice of prior distributions that we imposed in the fit, we repeated the MCMC analyses after relaxing or removing the Gaussian prior for $v\sin i$ or $T_c$. 
We found that while changing the Gaussian width of the prior to three times the original width did not significantly change $\lambda$ for both $v\sin i$ and $T_c$, completely removing the Gaussian prior (i.e., imposing a uniform prior) on $T_c$ led to a very poor constraint on the obliquity ($\lambda=3_{-124}^{+115}$ degrees). Assuming a uniform prior for $v\sin i$ resulted in an apparently consistent result ($\lambda=-12_{-58}^{+70}$ degrees), but we found a strong degeneracy in the fitting parameters ($v\sin i$ and $R_p/R_s$ in particular), and almost no meaningful constraints were obtained for $v\sin i$ and $R_p/R_s$.

\subsection{Doppler-Shadow Analysis} \label{sec:Doppler}

\begin{figure*}
\centering
\includegraphics[width=17cm]{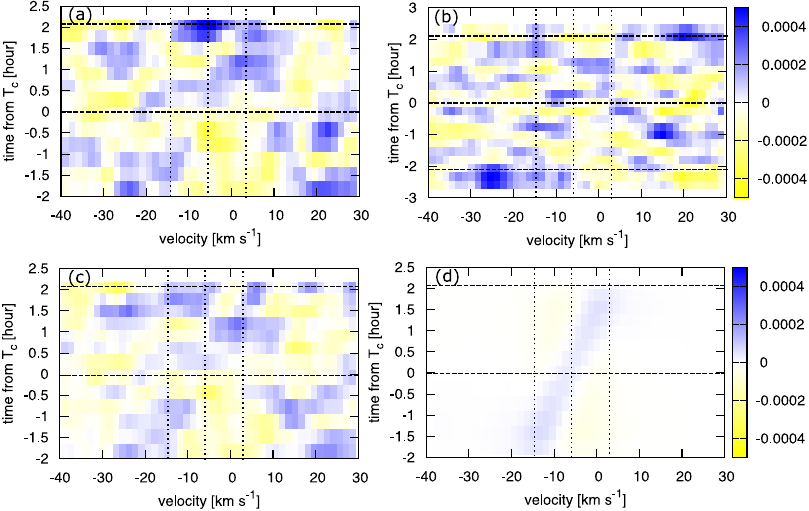}
\caption{
Residual CCF contours after subtracting the mean out-of-transit CCFs for
the 2020 (panel (a)) and 2021 (panel (b)) datasets. The color contour represents the fractional variation of the residual CCF. The horizontal dashed line 
show the transit center and times of ingress and egress. 
The vertical dotted lines correspond to the stellar-line center and approximate line
edges ($\pm v\sin i$ from the line center). 
The combined CCF contour using the two datasets is depicted in panel (c), and
its best-fit theoretical model is presented in panel (d). 
}
\label{fig:tomography}
\end{figure*}

Unfortunately, our measurements of the RM effect did not provide a good constraint
on the stellar obliquity, likely owing to the large RV uncertainties and degeneracy
in the system parameters; the transit impact parameter $b$ of K2-33b is relatively low, 
in which case the classical RM modeling is known to severely suffer
from the parameter degeneracy \citep[e.g.,][]{2011ApJ...738...50A}. 
We thus attempted a secondary analysis to constrain the obliquity based on 
the Doppler-shadow (or more often referred as ``Doppler tomography") technique
\citep[e.g.,][]{2010MNRAS.407..507C, 2017AJ....154..137J}.

The Doppler-shadow technique has already been applied to IRD data
to measure $\lambda$ 
\citep[e.g.,][]{2020ApJ...890L..27H, 2020ApJ...899L..13H, 2020MNRAS.498L.119G, 2022MNRAS.509.2969G}, 
and we follow the same approach here. Briefly, 
we computed CCFs for individual frames using GJ 436's spectrum as a CCF template
as described in Section \ref{sec:RM}, and derived the mean out-of-transit CCF after 
correcting for the barycentric motion of Earth. Since K2-33 exhibits very high 
stellar activity and two transits (2020 and 2021) were observed by IRD, we combined 
nightly frames on each transit night individually to focus on the relative line-profile
variations within the night. We then subtracted the mean CCF from the individual 
CCFs for each night to extract ``residual" CCF variations against time\footnote{Ideally, 
we should combine only out-of-transit frames to emphasize the transit-induced
CCF variations, but we were unable to do so due to insufficient out-of-transit
data (especially in 2020). We therefore combined all frames taken each night
to obtain the master CCF. }. 
For the 2020 data set, we discarded the first frame from the analysis 
due to the very low S/N ratio of the original spectrum. 
The top panels of Figure \ref{fig:tomography} display those residual CCF maps for the
2020 (panel (a)) and 2021 (panel (b)) transits. 
The three vertical dotted lines represent the CCF line-profile center (middle) and 
its $\pm 8.6$ km s$^{-1}$, which approximately corresponds to $v\sin i$ of K2-33.

The residual CCF maps are noisy and the planet shadow is hardly recognisable. 
We thus opted to combine the two transits to mitigate the impact of low S/N ratios
for each transit. 
Panel (c) of Figure \ref{fig:tomography} presents the combined (by weighted mean) 
residual CCFs using both 2020 and 2021 data sets. 
The map still suffers from systematic patterns, but it suggests a planet shadow
around the blueshifted edge of the line profile ($v\approx -15$ km s$^{-1}$) 
at ingress and that moves redward until the egress ($v\approx 2.5$ km s$^{-1}$).

To model the observed CCF residuals and estimate the system parameters, we synthesized mock IRD spectra that mimic in-transit
spectra during a transit of K2-33b. The details of the model calculation are 
given in \citet{2020ApJ...899L..13H}; we created a
total of 2142 mock spectra, changing $v\sin i$ (from $4.0$ to $12.0$ km s$^{-1}$) and 
the two-dimensional position of the planet on the stellar disc. 
Given the good agreement between
our estimate of $R_p/R_s$ from the classical RM modelling (Section \ref{sec:RM}) and 
that obtained by HST observations in the near-infrared ($1.088-1.68\,\mu$m), 
we fixed $R_p/R_s$ at 0.367 \citep{2023AJ....165...23T} generating mock transit spectra. 
All mock IRD spectra were then subject to the CCF analysis using GJ 436
as a template. With all these CCFs, one can produce a planet shadow model
in the residual CCF map for a given set of system parameters ($v\sin i$, $\lambda$, 
$a/R_s$, etc) by interpolations.

Following \citet{2020ApJ...899L..13H}, we performed an MCMC analysis to estimate 
the obliquity $\lambda$ as well as the other system parameters for K2-33b. 
In doing so, the time sequence of the  combined residual CCF (panel (c) of Figure \ref{fig:tomography}) 
was fitted by the above CCF model, and we derived the posterior distributions
for the fitting parameters: $v\sin i$, $\lambda$, $a/R_s$, $b$, and $T_c$. 
As in Section \ref{sec:RM}, we imposed Gaussian priors on $v\sin i$, $a/R_s$, 
$T_{14}$, and $T_c$. The other transit parameters, 
in particular $R_p/R_s$, $P$, $u_1$, and $u_2$, were held fixed to the literature and
theoretical values. The result of the fit is also given in Table \ref{hyo6}, and
the residual CCF model with the best-fit parameters is depicted in panel (d)
of Figure \ref{fig:tomography}. 
The spin-orbit angle $\lambda$ is now constrained with a better precision 
than that in the classical RM modelling, likely reflecting the discernible signal
of the planet shadow. Its estimation of $\lambda=-10_{-24}^{+22}$ degrees 
implies that K2-33b's orbit is almost aligned with the equatorial plane of the host star.

\section{He I Transmission Spectroscopy} 
\label{sec:HeI}

\begin{figure*}
\includegraphics[width=0.49\textwidth]{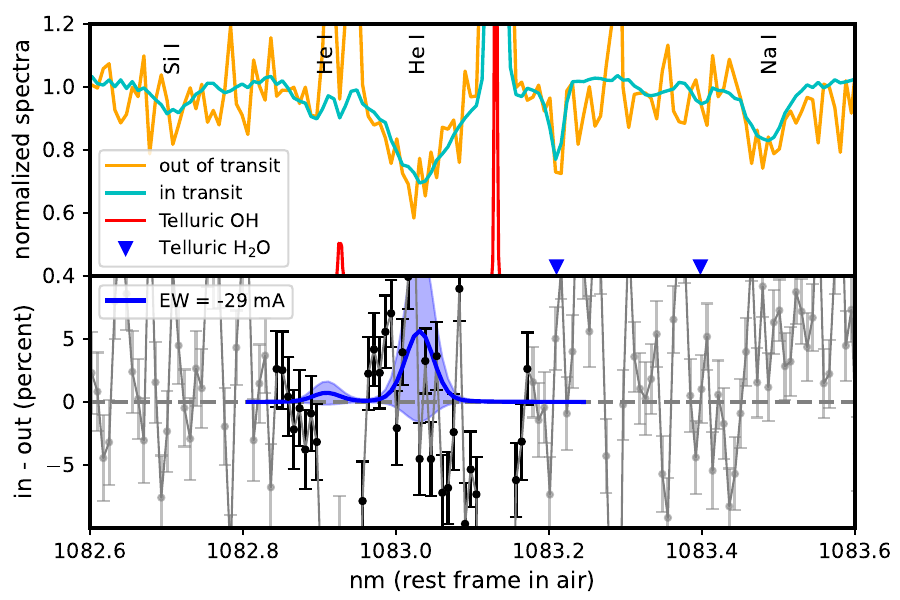}
\includegraphics[width=0.49\textwidth]{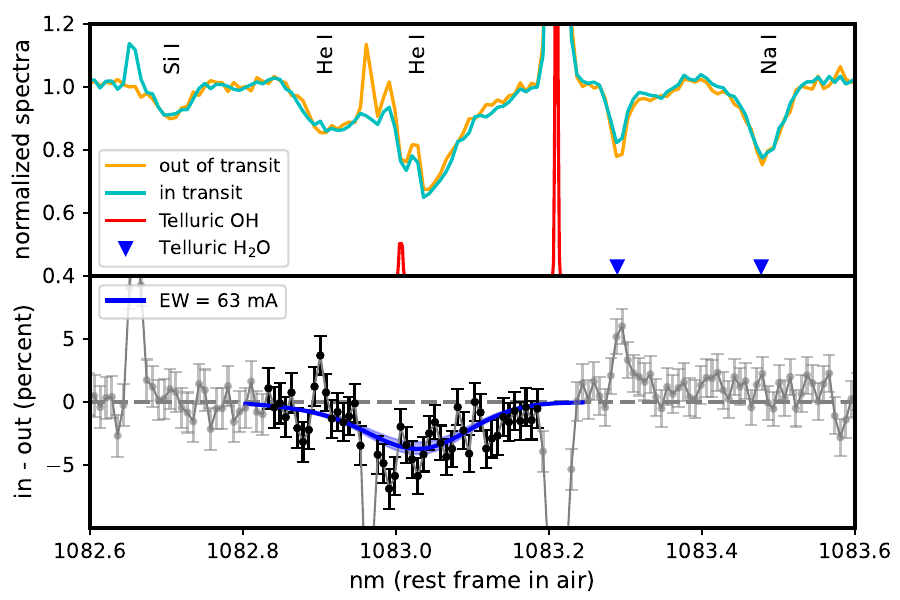}
    \caption{IRD spectra of K2-33 in the vicinity of the 1083.3 nm He~I triplet obtained during the transit opportunities on 8 June 2020 (left) and 24 April 2021 (right).  The upper panels show the in-transit (cyan) and out-of-transit (orange) spectra with significant stellar lines (including the He~I singlet and unresolved doublet) labeled, as well as telluric OH emission (red lines) and H$_2$O absorption (blue inverted triangles).  The lower panels contain the difference spectra (black lines) along with the best-fit He~I line (blue curves) and the 99\% upper limit (magenta lines), with the EWs reported in the legend.}
    \label{fig:hei_spec}
\end{figure*}

We constructed summed, normalised spectra of K2-33 inside and outside of each of the two transits of ``b" , and calculated difference spectra that should contain any absorption by the planet's extended atmosphere/wind (Fig. \ref{fig:hei_spec}).  During both events the He~I triplet is affected by the weaker, bluer of two neighboring OH lines (vertical red lines in upper panels of Fig. \ref{fig:hei_spec}).  The quality of the difference spectrum from the 2020 transit (lower panel of Fig. \ref{fig:hei_spec}a) is poor, likely due to the low signal-to-noise of the available out-of-transit spectra, and there is no indication of excess absorption.  However, we identify possible absorption during the 2021 transit (lower panel of Fig. \ref{fig:hei_spec}b).  

To quantify any transit-associated absorption and evaluate its significance, we fit a two-parameter model He~I triplet line profile to unaffected regions of the spectrum (black points in the lower panels of Fig. \ref{fig:hei_spec}).  The profile has a variable amplitude (expressed as an equivalent width, EW) and is broadened by the instrument resolution and gas temperature ($10^4$ K), plus variable Doppler broadening that represents the collective motion of the planet and its wind.  We determined 95\% confidence ratios by fitting 1000 Monte Carlo representations of the actual data with added random Gaussian noise.   The best-fit profiles for the 2020 and 2021 transits are shown as the heavy blue curves in the lower panels of Fig. \ref{fig:hei_spec} and have EWs of -29 m\AA{} and +63 m\AA{}, respectively.  The  95\% confidence range are plotted as the light blue shaded areas in Fig. \ref{fig:hei_spec} and are from -63 to +8 m\AA{} and and 55--71 m\AA{}, respectively. Thus while there is no significant feature in the 2020 transit, the anomalous absorption during the 2021 transit appears statistically significant.  However, these calculations do not include  systematic (time-correlated) and ``red" (wavelength correlated) noise and thus the significance of the 2021 detection should be interpreted with caution.

\begin{figure*}
    \centering
    \includegraphics[width=0.49\textwidth]{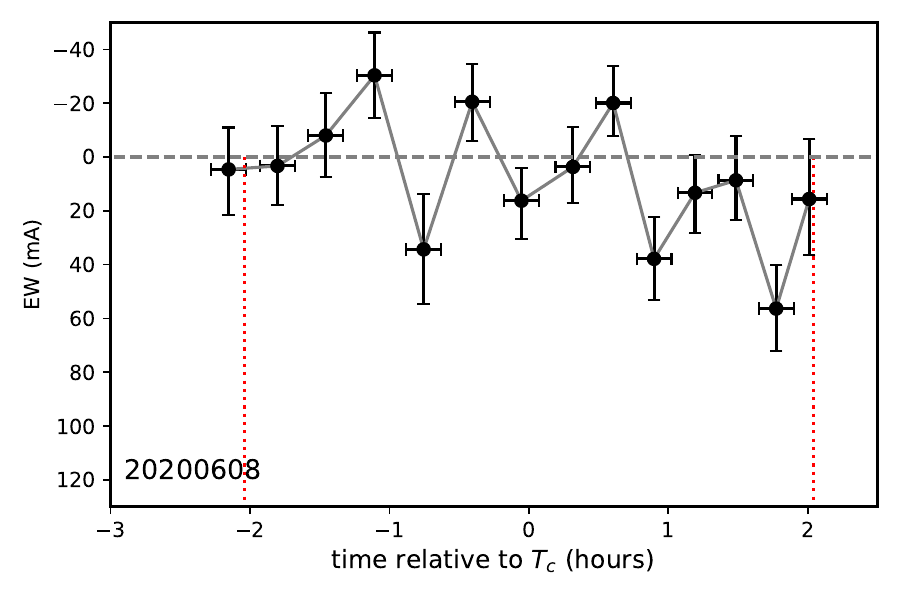}
    \includegraphics[width=0.49\textwidth]{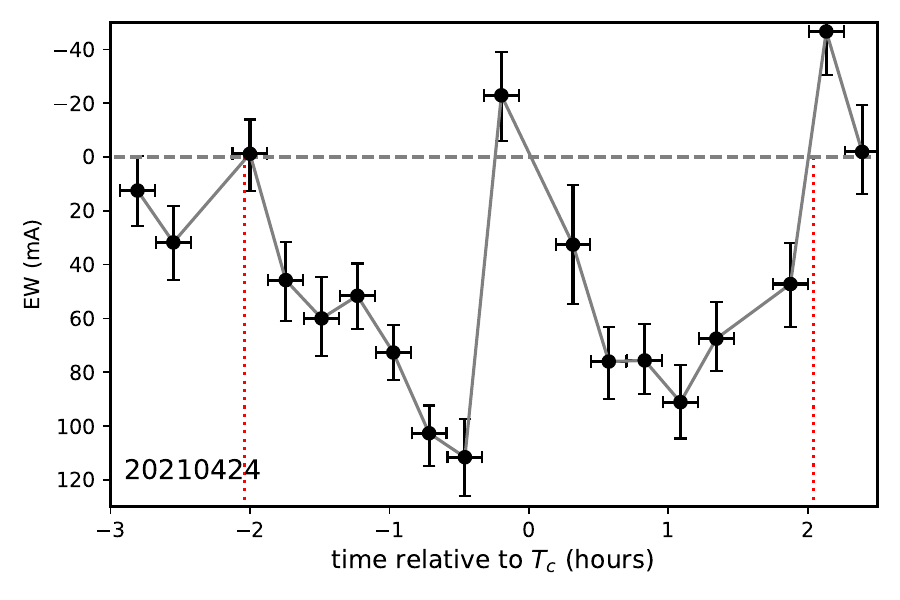}
    \caption{Time series of the He I line anomalous EW during the 2020 (left) and 2021 (right) transits of K2-33b.  Vertical dotted lines mark transit ingress and egress.  Vertical error-bars were generated from 1000 Monte Carlo representations of each spectrum, and the horizontal error-bars represent the integration times.}
    \label{fig:hei_lightcurve}
\end{figure*}

We further searched for anomalous absorption by performing fits of the same profile to each individual spectrum differenced by the mean out-of-transit spectrum, in these cases fixing the Doppler broadening term to the best-fit value (39 km~s$^{-1}$) obtained from the fit to the mean spectra of the 2021 transit.  Uncertainties in each EW are calculated using the same Monte Carlo scheme describe above.  The He~I EW time series (Fig. \ref{fig:hei_lightcurve}) show no significant variation during the 2020 transit ($\chi^2=31.0$ for 13 d.o.f.) but contains a clear transit-like trend corresponding to the predicted transit interval during the 2021 transit ($\chi^2=160.1$ for 17 d.o.f.).   

\begin{figure}
    \centering
    \includegraphics[width=\columnwidth]{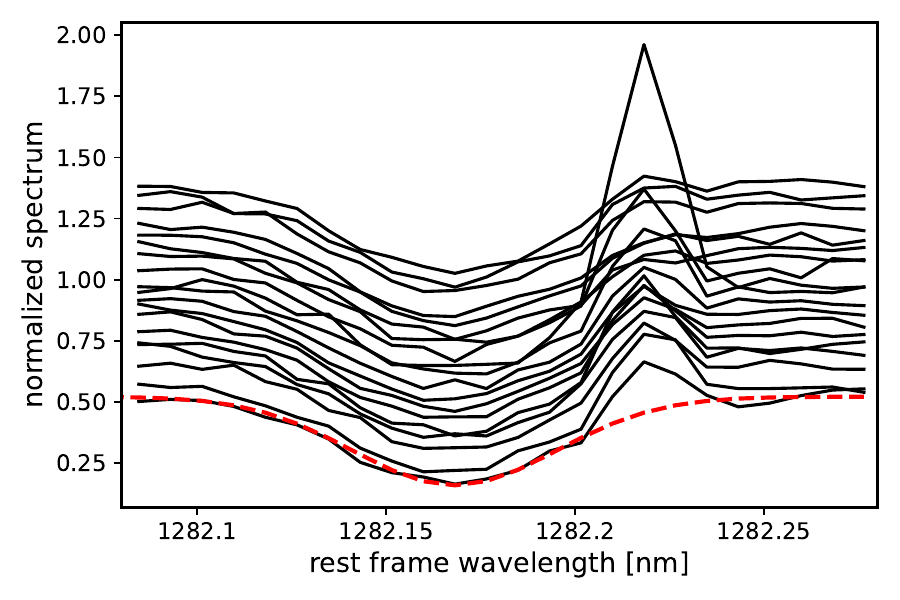}
    \caption{Time-series spectra of K2-33 around the 1.282 $\mu$m Paschen $\beta$ line of H~I during the 2021 transit of `b', showing the photospheric absorption line and the red-shifted emission line from a flare.  The spectra are temporally ordered from bottom to top and the dashed red line is the best-fit Voigt line profile used as a baseline to compute emission strength.}
    \label{fig:paschen_spec}
\end{figure}

\begin{figure}
    \centering
    \includegraphics[width=\columnwidth]{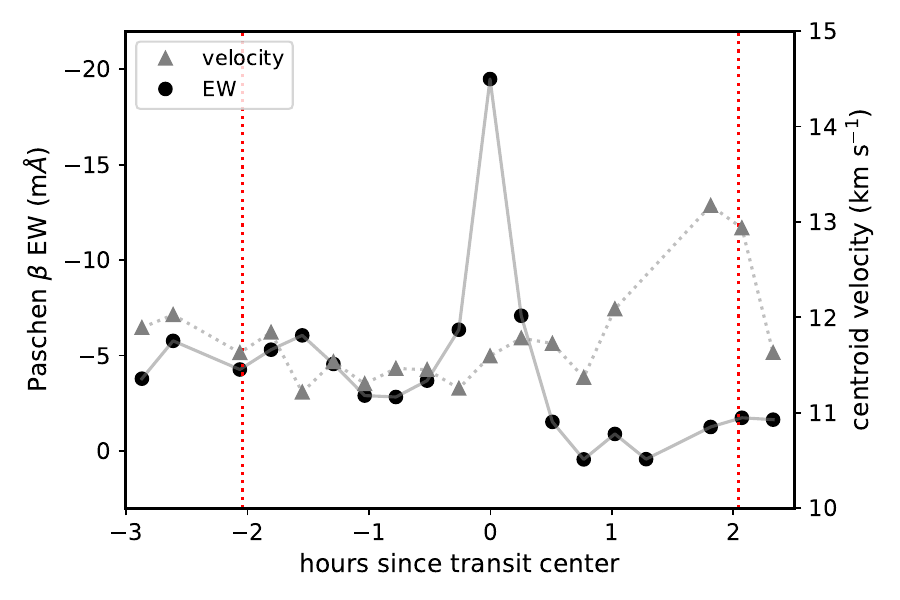}
    \caption{Variation in equivalent width and centroid red-shift (as a velocity) of emission in the line Paschen $\beta$ line from the flare.  The vertical dotted red lines are the ingress and egress times of the transit.}
    \label{fig:paschen_ew}
\end{figure}

The 2021 He~I time-series contains a large positive excursion near transit center.  We speculate that this is the result of a flare on this active star: strong flare-associated He~I emission has been observed on other active M dwarfs \citep{Schmidt2012,Hintz2020,Kanodia2022,HowardW2023} and the decay timescale ($\sim30$ min) is characteristic of flares.  Conclusive evidence for a flare comes from an analysis of the 1.2822 $\mu$m  Paschen $\beta$ line of H~I in the same spectra:  this reveals simultaneous, transient emission superposed on the photospheric absorption line (Figs. \ref{fig:paschen_spec} and \ref{fig:paschen_ew}).  The 11--12 km~sec$^{-1}$ red-shift is too large to be produced by stellar rotation ($V_{\rm eq} \approx 8$~km~s$^{-1}$) but might be produced by electron beaming \citep{Kowalski2022}.   Even more curious: in contrast to observations of emission in the Paschen lines during other flares \citep[e.g.,]{Kanodia2022,Fuhrmeister2023b}, this line is narrow and only marginally resolved (FWHM $\approx 0.03$ nm).  
We re-fit the in-transit spectrum excluding the two epochs most affected by the flare; the best-fit EW was unchanged (63 m\AA), but the 95\% confidence range shrank to 58--68 m\AA.

We used the best-fit and upper limit values of the EW from the 2021 transit to constrain possible mass loss from the planet, based on a model of an isothermal, spherically symmetric Parker wind with a solar H/He composition as described in \citet{2020MNRAS.495..650G}.   A key input to the model is the X-ray/EUV emission from the host star, which is responsible for formation of the triplet He~I state via ionization of He~I, but for which essentially nothing is known observationally.  In lieu of data, we adopted the panchromatic spectrum of GJ 729, an active, rapidly-rotating (2.85 day period) M3.5 dwarf (compared to the M3 spectral type of K2-33) from the Mega-MUSCLES survey planet-hosting K and M dwarfs \citep{2021ApJ...911...18W}.\footnote{We also repeated calculations using a spectrum of the older M2 field dwarf GJ 832 from \citet{Fontenla2016} and found qualitatively similar results.}  Fluxes in specific X-ray and UV bands were adjusted using the empirical relations between these bands and Lyman-$\alpha$ emission developed by \citet{Linsky2014}.  The Lyman-$\alpha$ emission was in turn estimated from the star's X-ray luminosity and the empirical relation of \citet{Linsky2013}.  In turn the X-ray luminosity was estimated using the star's Rossby number and the relation of \citet{Wright2018}.  (K2-33 does not have a corresponding source in the Second \emph{ROSAT} All Sky Survey source catalog \citep{Boller2016}, but the flux at Earth based on the estimated X-ray luminosity is only $2 \times 10^{-13}$ ergs s$^{-1}$ cm$^{-2}$, comparable to the detection limit of the survey.) The Rossby number $P/\tau_c$ was calculated to be 0.10 (in the saturated regime) using the rotation period $P$ from \citet{Klein2020}  and a convective turnover time $\tau_c$ of 65 days calculated using the stellar luminosity of \citet{2023AJ....165...23T} and scaling by the square-root of the luminosity \citep{Pizzolato2003} relative to a solar value of 25 days.  We assumed two values for the planet mass (which sets the atmospheric scale height and gravitational potential well but is a poorly constrained parameter of the wind model) of 25 and 5 M$_{\oplus}$.  The first is motivated by the empirical relation between mass and radius of $M_p \sim R_p^2$, while the second invokes an ``extreme" scenario of a super-Earth rocky core surrounded by about an Earth-mass of H and He.

\begin{figure*}

\includegraphics[width=0.49\textwidth]{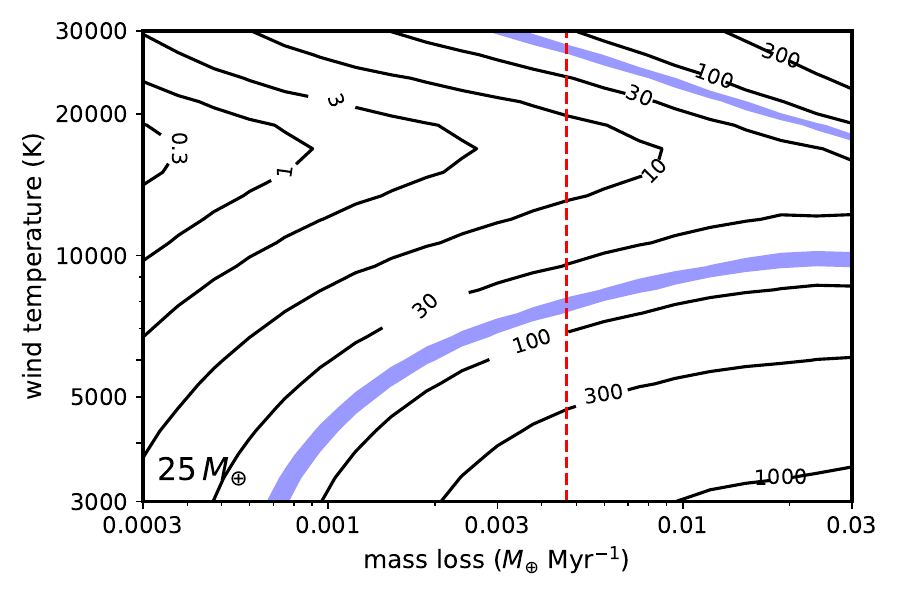}
\includegraphics[width=0.49\textwidth]{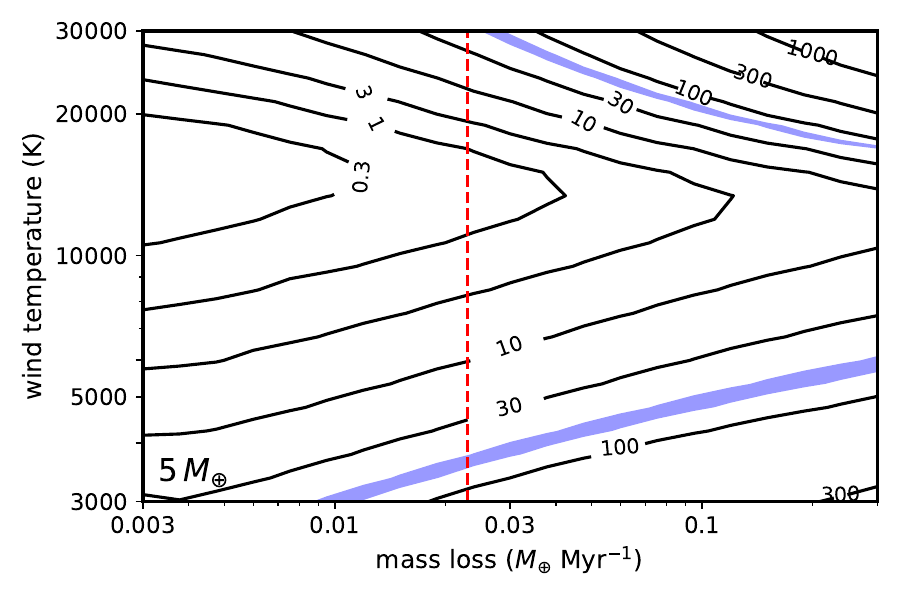}
    \caption{Predicted absorption in the He~I from a spherical, isothermal Parker wind model of an escaping H/He atmosphere around K2-33b, assuming two different planet masses (25 and 5$M_{\oplus}$).  Contours of EW are in m\AA.  The blue region is the absorption inferred from observations of the 2021 transit (Fig. \ref{fig:hei_spec}b).   Note the horizontal axes are different between the left and right panels.  Vertical dashed red lines mark the estimated XUV energy-limited escape rate assuming 100\% efficiency.}
    \label{fig:hei_model}
\end{figure*}

In Fig. \ref{fig:hei_model}, calculated equivalent widths in m\AA{} are plotted as a function of mass-loss rate (between 0.0003-0.3 M$_{\oplus}$ Myr$^{-1}$, bracketing the estimated energy-limited escape rates, see below) and wind temperature (3000-30000K).  The blue regions are the 95\% confidence range for anomalous absorption in the 2021 transit (Fig. \ref{fig:hei_spec}b).  A lower planetary mass permits higher escape rates, but for a given escape rate the He~I column density along the line of sight is lower due to slower wind speeds.  The total incident X-ray plus EUV and Ly-$\alpha$ flux (averaged over the planet) is predicted to be $1.5 \times 10^4$ ergs s$^{-1}$ cm$^{-2}$, and the corresponding XUV energy-limited escape rates (for 100\% efficiency) are 0.0047 and 0.023 M$_{\oplus}$~Myr$^{-1}$ for 25 and 5 M$_{\oplus}$ (vertical red dashed lines in Fig. \ref{fig:hei_model}).  For the 5 M$_{\oplus}$ case, the available XUV emission is insufficient to induce a wind of sufficient density to produce the observed anomaly.\footnote{Given the young age of K2-33b, this does not excluded some other more transient process such as core-powered mass loss or a giant impact.}   In the 25 M$_{\oplus}$ case, and XUV-driven wind can plausibly produce the anomaly but only if the wind temperature is low and the escape efficiency is high, or the stellar XUV luminosity is significantly higher than predicted.


\section{Discussion} \label{sec:discussion}


Our measurements suggest a low stellar obliquity for K2-33.  Our constraint on $\lambda$ ($-10_{-24}^{+22}$ deg) can be combined with the stellar inclination, which is the obliquity projected onto the line-of-sight, to constrain the unprojected obliquity.  The stellar inclination is estimated via $v\sin i$ and the equatorial rotation velocity 
derived from the rotation period and radius of the star. \citet{2016AJ....152...61M} constrained the stellar inclination angle of K2-33 to $i_s>63^\circ$ at $1\,\sigma$;  here we revised this constraint based on the updated rotation velocity of $v\sin i=8.6\pm 0.4$ km s$^{-1}$.  Adopting the rotation period of $P_\mathrm{rot}=6.29\pm 0.17$ days \citep{2016AJ....152...61M} and making use of the Bayesian inference scheme described by \citet{2020AJ....159...81M}, we computed the posterior distribution of $\cos i_s$.  This yielded $\cos i_s=0.20_{-0.06}^{+0.17}$, corresponding to  $i_s=78_{-9}^{+8}$ deg.  Thus, both the sky-projected and line-of-sight stellar obliquities of K2-33 are low.  Together with $\lambda=-10_{-24}^{+22}$ degree (Section \ref{sec:Doppler}) and the orbital inclination of $i_o=89.1_{-1.1}^{+0.6}$ deg \citep{2016AJ....152...61M} and using the geometric equation for the stellar obliquity \citep[e.g., Eq. (9) of ][]{2009ApJ...696.1230F},  we further derived the 3-dimensional (unprojected) stellar obliquity as $\psi=23_{-11}^{+16}$ deg. 


\begin{figure}
\centering
\includegraphics[width=8.5cm]{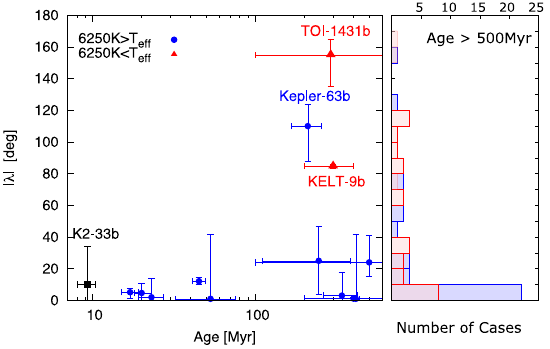}
\caption{Observed projected obliquity $|\lambda|$ as a function of system age ($<500$ Myr). Planets around hot stars ($T_\mathrm{eff}>6250$ K) are plotted by red triangle, while those around cool stars ($T_\mathrm{eff}<6250$ K) are shown in blue. K2-33b (black square) is the youngest planet with a measured obliquity. 
The right panel shows the histograms of $|\lambda|$ for systems older than $500$ Myr. Those for hot and cool stars are shown in red and blue, respectively. Data downloaded from the NASA Exoplanet Archive (\url{https://exoplanetarchive.ipac.caltech.edu/index.html}).
}
\label{fig:age-lambda}
\end{figure}

The low stellar obliquity of the K2-33 system is in line with the results of other obliquity measurements for very young transiting planets;
To this point, stellar obliquities for six transiting systems younger than $100$ Myr
(AU Mic, HIP 67522, V1298 Tau, DS Tuc A, TOI-942, and K2-33) have been reported 
\citep{2020ApJ...899L..13H, 2020A&A...643A..25P, 2020A&A...641L...1M, 
2021AJ....162..137A, 2021ApJ...922L...1H, 
2022MNRAS.509.2969G, 2022AJ....163..247J, 2020ApJ...892L..21Z,
2020AJ....159..112M, 2021ApJ...917L..34W}, 
and all measurements indicated
low obliquities that are compatible with spin-orbit alignment. 
In other words, no evidence for a primordial spin-orbit misalignment was found 
for newborn planets. 
Since the timescale of obliquity damping by tides is 
generally longer than $\sim 1$ Gyr \citep{2012ApJ...757...18A, 2022PASP..134h2001A}, those planets ($<1$ Gyr) were born 
with a primordially aligned orbit with the stellar equatorial plane, evidence that 
they 
formed at their present location or migrated inwards along the original disc plane
\citep[e.g.,][]{1996Natur.380..606L, 2008ApJ...673..487I, 2014prpl.conf..667B} 
without chaotic orbital disturbance. 
This is in marked contrast to the distribution of $\lambda$ for older counterparts
($\gtrsim1$ Gyr), a significant fraction of which is known to exhibit spin-orbit misalignment (Fig. \ref{fig:age-lambda}) \citep[also see e.g.,][]{2022PASP..134h2001A}. 
However, one should notice that 
many of these cases are isolated hot Jupiters while young planets with obliquity measurements are generally smaller in size or have longer orbital periods; those measurements are possibly probing different populations of close-in planets. 
Note that when we extend the age range to $100-1000$ Myr, 
we find a fraction of planets having highly misaligned orbits
(Fig. \ref{fig:age-lambda}), but most of those are very hot jovian 
planets orbiting early-type stars with $T_\mathrm{eff}>6250$ K, 
a boundary that divides the mostly aligned and misaligned 
systems \citep{2010ApJ...718L.145W,2012ApJ...757...18A}. 
Further observations for planets with differing ages and sizes would help
us unveil the age evolution of the obliquity distribution. 


Although our 2021 observations of the K2-33 system show transit-associated excess absorption in the 1.083 $\mu$m of metastable ortho-He I consistent with an extended, escaping atmosphere, the significance of this detection awaits a better understanding of systematic effects, particularly changes in the stellar chromosphere line on hours-long timescales \citep[see also ][]{2022MNRAS.509.2969G}.   Although we lack sufficient baseline observations to definitely rule out a stellar source for the variation, the absence of significant variation during the 2020 transit (Fig. \ref{fig:hei_lightcurve}a) is encouraging.  As another assessment, we used other spectra of K2-33 taken outside of transit for RV monitoring (Hirano et al. in prep), to generate residuals between pairs of spectra obtained at least 30 min and at most 5 hours apart, and performed fits of the same He~I triplet profile to retrieve spurious EWs.  We found that, with the exception of spectra from a single run in 2019 that appears to be affected by systematics, the scatter in best-fit EWs fell below that determined from the 2021 transit. 


The lack of a detectable He~I anomaly in 2020 then suggests that the planetary wind varies on year-long timescales, e.g., with the changing level of magnetic activity on and XUV irradiation from the star.  Magnetic cycling of a few years period has been observed on very cool dwarfs stars, including T Tauri stars \citep[e.g.,][]{Klein2021,Finociety2023,LinC2023}.  Ultimately, there is no substitute for additional transit observations with adequate out-of-transit baseline to resolve the issue.  Higher signal-to-noise observations that can avoid or better correct contamination by tellurics, especially the intense OH lines, are also needed.   Finally, observations and analysis of K2-33 in X-rays, e.g. with \emph{eROSITA} and \emph{XMM-Newton}, are needed to better estimate the high-energy irradiance of its planet.\footnote{K2-33 was detected in an unpublished 33-ksec observation by \emph{XMM-Newton} (ID, 0824840101, PI M. Salz). \emph{HST}obtained 1150-1700\,\AA\ spectra of K2-33, including the Ly-$\alpha$ line, with STIS Cycle 24 Proposal 14887, PI B. Benneke), but due to the distance and interstellar extinction, the S/N ratio is too low (K. France, pers. comm.).}   If the atmosphere of K2-33b is escaping at a rate of $3 \times 10^{-3}$ $M_{\oplus}$~Myr$^{-1}$ (Fig. \ref{fig:hei_model}), after several hundred Myr most or all of a primordial H/He envelope would be lost, leaving this sub-Neptune as a future super-Earth.


The reduced radius in the near infrared revealed by \citet{2023AJ....165...23T} was not confirmed 
for our RV data due to the large RV uncertainty and/or degeneracy 
in the fitting parameters. To see what extent we can constrain $R_p/R_s$ in case
of a perfect spin-orbit alignment, which is supported by the Doppler-shadow
analysis (Sec. \ref{sec:Doppler}), we repeated the MCMC fit to the RV data assuming $\lambda=0$ degree (while $v\sin i$ was allowed to vary with the same Gaussian prior). 
As a result, we found $R_p/R_s=0.035_{-0.012}^{+0.009}$, which gives a slightly
tighter constraint (i.e., deviated from the {\it K2} value by $>1\,\sigma$), but
the smaller radius in the IRD wavelength range is still inconclusive. 

To further discuss the atmospheric escape/evolution of K2-33b as well as the origin of the smaller near-infrared radius, constraining its planet mass would be an urgent task. 
Near-infrared observations such as by IRD have a marked advantage over visible RV
measurements in that activity-induced RV variations are known to be mitigated by virtue of improved spots' contrasts at longer wavelengths \citep[e.g.,][]{2012ApJ...761..164C, 2021AJ....162..104M}. 
However, the detection of a large flare in the IRD data during a mere 10 hr of monitoring suggests that such events are frequent, and less energetic events may be even more frequent.  Flares are a source of astrophysical noise or ``jitter" in RV measurements \citep[e.g.,][]{Reiners2009,Pietrow2023} and indeed, we see our RV measurements contemporaneous with the flare (see an RV excursion around $T_c$ in Fig. \ref{fig:RM}b).  Flare may be an impediment to precise infrared RVs for young, active stars such as the host of K2-33b, but their identification using ``fingerprint" lines such as Paschen $\beta$ offers a potential way forward.



\section{Conclusions}\label{sec:conlusions}


We obtained high-resolution infrared ($YJH$-band) spectroscopy of the 8-11 Myr-old low-mass star K2-33 during two transits of its close-in $5R_{\oplus}$ planet, one of the youngest such objects known. By modelling the RV variations during the transits, we found hints of low projected stellar obliquity for K2-33 ($\lambda=-6_{-58}^{+61}$ deg) and smaller radius of K2-33b in the near-infrared band ($R_p/R_s=0.037_{-0.017}^{+0.013}$) than in the optical, but those findings are inconclusive due to the lower RV precision.  We also constrained the obliquity by analysing the ``shadow" of the planet in cross-correlation profiles with wavelength to better constrain the obliquity ($\lambda=-10_{-24}^{+22}$ deg).   Combining the latter constraint with the orbital and stellar inclination angles revealed by the transit and $v\sin i$ measurements, we found a low de-projected spin-orbit angle of $\psi=23_{-11}^{+16}$ deg for K2-33b. 

We also compared the mean spectra inside and outside of transits of K2-33b to search for excess absorption in the 1083 nm triplet of metastable ortho-He I. While the 2020 data contain no evidence for anomalous absorption, the 2021 transit shows possible absorption consistent with an extended atmosphere, although we cannot definitely rule out an astrophysical signal from the star.  Our modeling of the signal with a spherical model of a planetary wind suggests mass loss of at most a few Earth-mass of solar-composition H/He over a Gyr, potentially driven by XUV irradiation, and dependent on the (as yet determined) planet mass.

\section*{Acknowledgements}

We thank the reviewer, Jiayin Dong, for the insightful comments to improve the manuscript. 
We are grateful to Kazumasa Ohno for helping us derive the revised ephemeris for the target. This work was supported by JSPS KAKENHI Grant Numbers JP19K14783, JP21H00035, JP23H01227, JP18H05442. The data analysis was carried out, in part, on the Multi-wavelength Data Analysis System operated by the Astronomy Data Center (ADC), National Astronomical Observatory of Japan.  EG was supported by NASA Exoplanets Research Program Awards 80NSSC20K0957 and 80NSSC20K0251.

\section*{Data Availability}

The original and reduced data used in this paper can be made available upon requests.




\bibliographystyle{mnras}

\begin{thebibliography}{}
\makeatletter
\relax
\def\mn@urlcharsother{\let\do\@makeother \do\$\do\&\do\#\do\^\do\_\do\%\do\~}
\def\mn@doi{\begingroup\mn@urlcharsother \@ifnextchar [ {\mn@doi@}
  {\mn@doi@[]}}
\def\mn@doi@[#1]#2{\def\@tempa{#1}\ifx\@tempa\@empty \href
  {http://dx.doi.org/#2} {doi:#2}\else \href {http://dx.doi.org/#2} {#1}\fi
  \endgroup}
\def\mn@eprint#1#2{\mn@eprint@#1:#2::\@nil}
\def\mn@eprint@arXiv#1{\href {http://arxiv.org/abs/#1} {{\tt arXiv:#1}}}
\def\mn@eprint@dblp#1{\href {http://dblp.uni-trier.de/rec/bibtex/#1.xml}
  {dblp:#1}}
\def\mn@eprint@#1:#2:#3:#4\@nil{\def\@tempa {#1}\def\@tempb {#2}\def\@tempc
  {#3}\ifx \@tempc \@empty \let \@tempc \@tempb \let \@tempb \@tempa \fi \ifx
  \@tempb \@empty \def\@tempb {arXiv}\fi \@ifundefined
  {mn@eprint@\@tempb}{\@tempb:\@tempc}{\expandafter \expandafter \csname
  mn@eprint@\@tempb\endcsname \expandafter{\@tempc}}}

\bibitem[\protect\citeauthoryear{{Addison} et~al.,}{{Addison}
  et~al.}{2021}]{2021AJ....162..137A}
{Addison} B.~C.,  et~al., 2021, \mn@doi [\aj] {10.3847/1538-3881/ac1685}, \href
  {https://ui.adsabs.harvard.edu/abs/2021AJ....162..137A} {162, 137}

\bibitem[\protect\citeauthoryear{{Albrecht} et~al.,}{{Albrecht}
  et~al.}{2011}]{2011ApJ...738...50A}
{Albrecht} S.,  et~al., 2011, \mn@doi [\apj] {10.1088/0004-637X/738/1/50},
  \href {https://ui.adsabs.harvard.edu/abs/2011ApJ...738...50A} {738, 50}

\bibitem[\protect\citeauthoryear{{Albrecht} et~al.,}{{Albrecht}
  et~al.}{2012}]{2012ApJ...757...18A}
{Albrecht} S.,  et~al., 2012, \mn@doi [\apj] {10.1088/0004-637X/757/1/18},
  \href {https://ui.adsabs.harvard.edu/abs/2012ApJ...757...18A} {757, 18}

\bibitem[\protect\citeauthoryear{{Albrecht}, {Dawson}  \& {Winn}}{{Albrecht}
  et~al.}{2022}]{2022PASP..134h2001A}
{Albrecht} S.~H.,  {Dawson} R.~I.,   {Winn} J.~N.,  2022, \mn@doi [\pasp]
  {10.1088/1538-3873/ac6c09}, \href
  {https://ui.adsabs.harvard.edu/abs/2022PASP..134h2001A} {134, 082001}

\bibitem[\protect\citeauthoryear{{Baruteau} et~al.,}{{Baruteau}
  et~al.}{2014}]{2014prpl.conf..667B}
{Baruteau} C.,  et~al., 2014, in {Beuther} H.,  {Klessen} R.~S.,  {Dullemond}
  C.~P.,   {Henning} T.,  eds, Protostars and Planets VI. pp 667--689
  (\mn@eprint {arXiv} {1312.4293}),
  \mn@doi{10.2458/azu_uapress_9780816531240-ch029}

\bibitem[\protect\citeauthoryear{{Boller}, {Freyberg}, {Tr{\"u}mper}, {Haberl},
  {Voges}  \& {Nandra}}{{Boller} et~al.}{2016}]{Boller2016}
{Boller} T.,  {Freyberg} M.~J.,  {Tr{\"u}mper} J.,  {Haberl} F.,  {Voges} W.,
  {Nandra} K.,  2016, \mn@doi [\aap] {10.1051/0004-6361/201525648}, \href
  {https://ui.adsabs.harvard.edu/abs/2016A&A...588A.103B} {588, A103}

\bibitem[\protect\citeauthoryear{{Bourrier} et~al.,}{{Bourrier}
  et~al.}{2018}]{2018Natur.553..477B}
{Bourrier} V.,  et~al., 2018, \mn@doi [\nat] {10.1038/nature24677}, \href
  {https://ui.adsabs.harvard.edu/abs/2018Natur.553..477B} {553, 477}

\bibitem[\protect\citeauthoryear{{Claret}, {Hauschildt}  \& {Witte}}{{Claret}
  et~al.}{2013}]{2013A&A...552A..16C}
{Claret} A.,  {Hauschildt} P.~H.,   {Witte} S.,  2013, \mn@doi [\aap]
  {10.1051/0004-6361/201220942}, \href
  {http://adsabs.harvard.edu/abs/2013A%26A...552A..16C} {552, A16}

\bibitem[\protect\citeauthoryear{{Collier Cameron} et~al.,}{{Collier Cameron}
  et~al.}{2010}]{2010MNRAS.407..507C}
{Collier Cameron} A.,  et~al., 2010, \mn@doi [\mnras]
  {10.1111/j.1365-2966.2010.16922.x}, \href
  {https://ui.adsabs.harvard.edu/abs/2010MNRAS.407..507C} {407, 507}

\bibitem[\protect\citeauthoryear{{Crockett}, {Mahmud}, {Prato}, {Johns-Krull},
  {Jaffe}, {Hartigan}  \& {Beichman}}{{Crockett}
  et~al.}{2012}]{2012ApJ...761..164C}
{Crockett} C.~J.,  {Mahmud} N.~I.,  {Prato} L.,  {Johns-Krull} C.~M.,  {Jaffe}
  D.~T.,  {Hartigan} P.~M.,   {Beichman} C.~A.,  2012, \mn@doi [\apj]
  {10.1088/0004-637X/761/2/164}, \href
  {https://ui.adsabs.harvard.edu/abs/2012ApJ...761..164C} {761, 164}

\bibitem[\protect\citeauthoryear{{David} et~al.,}{{David}
  et~al.}{2016}]{2016Natur.534..658D}
{David} T.~J.,  et~al., 2016, \mn@doi [\nat] {10.1038/nature18293}, \href
  {https://ui.adsabs.harvard.edu/abs/2016Natur.534..658D} {534, 658}

\bibitem[\protect\citeauthoryear{{David}, {Petigura}, {Luger},
  {Foreman-Mackey}, {Livingston}, {Mamajek}  \& {Hillenbrand}}{{David}
  et~al.}{2019}]{2019ApJ...885L..12D}
{David} T.~J.,  {Petigura} E.~A.,  {Luger} R.,  {Foreman-Mackey} D.,
  {Livingston} J.~H.,  {Mamajek} E.~E.,   {Hillenbrand} L.~A.,  2019, \mn@doi
  [\apjl] {10.3847/2041-8213/ab4c99}, \href
  {https://ui.adsabs.harvard.edu/abs/2019ApJ...885L..12D} {885, L12}

\bibitem[\protect\citeauthoryear{{Fabrycky} \& {Tremaine}}{{Fabrycky} \&
  {Tremaine}}{2007}]{2007ApJ...669.1298F}
{Fabrycky} D.,  {Tremaine} S.,  2007, \mn@doi [\apj] {10.1086/521702}, \href
  {https://ui.adsabs.harvard.edu/abs/2007ApJ...669.1298F} {669, 1298}

\bibitem[\protect\citeauthoryear{{Fabrycky} \& {Winn}}{{Fabrycky} \&
  {Winn}}{2009}]{2009ApJ...696.1230F}
{Fabrycky} D.~C.,  {Winn} J.~N.,  2009, \mn@doi [\apj]
  {10.1088/0004-637X/696/2/1230}, \href
  {https://ui.adsabs.harvard.edu/abs/2009ApJ...696.1230F} {696, 1230}

\bibitem[\protect\citeauthoryear{{Finociety} et~al.,}{{Finociety}
  et~al.}{2023}]{Finociety2023}
{Finociety} B.,  et~al., 2023, \mn@doi [\mnras] {10.1093/mnras/stad3012}, \href
  {https://ui.adsabs.harvard.edu/abs/2023MNRAS.526.4627F} {526, 4627}

\bibitem[\protect\citeauthoryear{{Fontenla}, {Linsky}, {Garrison}, {France},
  {Buccino}, {Mauas}, {Vieytes}  \& {Walkowicz}}{{Fontenla}
  et~al.}{2016}]{Fontenla2016}
{Fontenla} J.~M.,  {Linsky} J.~L.,  {Garrison} J.,  {France} K.,  {Buccino} A.,
   {Mauas} P.,  {Vieytes} M.,   {Walkowicz} L.~M.,  2016, \mn@doi [\apj]
  {10.3847/0004-637X/830/2/154}, \href
  {https://ui.adsabs.harvard.edu/abs/2016ApJ...830..154F} {830, 154}

\bibitem[\protect\citeauthoryear{{Fuhrmeister} et~al.,}{{Fuhrmeister}
  et~al.}{2023}]{Fuhrmeister2023b}
{Fuhrmeister} B.,  et~al., 2023, \mn@doi [\aap] {10.1051/0004-6361/202347161},
  \href {https://ui.adsabs.harvard.edu/abs/2023A&A...678A...1F} {678, A1}

\bibitem[\protect\citeauthoryear{{Fulton} et~al.,}{{Fulton}
  et~al.}{2017}]{2017AJ....154..109F}
{Fulton} B.~J.,  et~al., 2017, \mn@doi [\aj] {10.3847/1538-3881/aa80eb}, \href
  {http://adsabs.harvard.edu/abs/2017AJ....154..109F} {154, 109}

\bibitem[\protect\citeauthoryear{{Gaidos} et~al.,}{{Gaidos}
  et~al.}{2020a}]{2020MNRAS.495..650G}
{Gaidos} E.,  et~al., 2020a, \mn@doi [\mnras] {10.1093/mnras/staa918}, \href
  {https://ui.adsabs.harvard.edu/abs/2020MNRAS.495..650G} {495, 650}

\bibitem[\protect\citeauthoryear{{Gaidos} et~al.,}{{Gaidos}
  et~al.}{2020b}]{2020MNRAS.498L.119G}
{Gaidos} E.,  et~al., 2020b, \mn@doi [\mnras] {10.1093/mnrasl/slaa136}, \href
  {https://ui.adsabs.harvard.edu/abs/2020MNRAS.498L.119G} {498, L119}

\bibitem[\protect\citeauthoryear{{Gaidos} et~al.,}{{Gaidos}
  et~al.}{2022}]{2022MNRAS.509.2969G}
{Gaidos} E.,  et~al., 2022, \mn@doi [\mnras] {10.1093/mnras/stab3107}, \href
  {https://ui.adsabs.harvard.edu/abs/2022MNRAS.509.2969G} {509, 2969}

\bibitem[\protect\citeauthoryear{{Gaidos} et~al.,}{{Gaidos}
  et~al.}{2023}]{Gaidos2023}
{Gaidos} E.,  et~al., 2023, \mn@doi [\mnras] {10.1093/mnras/stac3301}, \href
  {https://ui.adsabs.harvard.edu/abs/2023MNRAS.518.3777G} {518, 3777}

\bibitem[\protect\citeauthoryear{{Ginzburg}, {Schlichting}  \&
  {Sari}}{{Ginzburg} et~al.}{2018}]{2018MNRAS.476..759G}
{Ginzburg} S.,  {Schlichting} H.~E.,   {Sari} R.,  2018, \mn@doi [\mnras]
  {10.1093/mnras/sty290}, \href
  {https://ui.adsabs.harvard.edu/abs/2018MNRAS.476..759G} {476, 759}

\bibitem[\protect\citeauthoryear{{Gupta}, {Nicholson}  \&
  {Schlichting}}{{Gupta} et~al.}{2022}]{2022MNRAS.516.4585G}
{Gupta} A.,  {Nicholson} L.,   {Schlichting} H.~E.,  2022, \mn@doi [\mnras]
  {10.1093/mnras/stac2488}, \href
  {https://ui.adsabs.harvard.edu/abs/2022MNRAS.516.4585G} {516, 4585}

\bibitem[\protect\citeauthoryear{{Heitzmann} et~al.,}{{Heitzmann}
  et~al.}{2021}]{2021ApJ...922L...1H}
{Heitzmann} A.,  et~al., 2021, \mn@doi [\apjl] {10.3847/2041-8213/ac3485},
  \href {https://ui.adsabs.harvard.edu/abs/2021ApJ...922L...1H} {922, L1}

\bibitem[\protect\citeauthoryear{{Hintz} et~al.,}{{Hintz}
  et~al.}{2020}]{Hintz2020}
{Hintz} D.,  et~al., 2020, \mn@doi [\aap] {10.1051/0004-6361/202037596}, \href
  {https://ui.adsabs.harvard.edu/abs/2020A&A...638A.115H} {638, A115}

\bibitem[\protect\citeauthoryear{{Hirano}, {Suto}, {Winn}, {Taruya}, {Narita},
  {Albrecht}  \& {Sato}}{{Hirano} et~al.}{2011}]{2011ApJ...742...69H}
{Hirano} T.,  {Suto} Y.,  {Winn} J.~N.,  {Taruya} A.,  {Narita} N.,  {Albrecht}
  S.,   {Sato} B.,  2011, \mn@doi [\apj] {10.1088/0004-637X/742/2/69}, \href
  {https://ui.adsabs.harvard.edu/abs/2011ApJ...742...69H} {742, 69}

\bibitem[\protect\citeauthoryear{{Hirano} et~al.,}{{Hirano}
  et~al.}{2016}]{2016ApJ...825...53H}
{Hirano} T.,  et~al., 2016, \mn@doi [\apj] {10.3847/0004-637X/825/1/53}, \href
  {https://ui.adsabs.harvard.edu/abs/2016ApJ...825...53H} {825, 53}

\bibitem[\protect\citeauthoryear{{Hirano} et~al.,}{{Hirano}
  et~al.}{2018}]{2018AJ....155..127H}
{Hirano} T.,  et~al., 2018, \mn@doi [\aj] {10.3847/1538-3881/aaa9c1}, \href
  {https://ui.adsabs.harvard.edu/abs/2018AJ....155..127H} {155, 127}

\bibitem[\protect\citeauthoryear{{Hirano} et~al.,}{{Hirano}
  et~al.}{2020a}]{2020PASJ...72...93H}
{Hirano} T.,  et~al., 2020a, \mn@doi [\pasj] {10.1093/pasj/psaa085}, \href
  {https://ui.adsabs.harvard.edu/abs/2020PASJ...72...93H} {72, 93}

\bibitem[\protect\citeauthoryear{{Hirano} et~al.,}{{Hirano}
  et~al.}{2020b}]{2020ApJ...890L..27H}
{Hirano} T.,  et~al., 2020b, \mn@doi [\apjl] {10.3847/2041-8213/ab74dc}, \href
  {https://ui.adsabs.harvard.edu/abs/2020ApJ...890L..27H} {890, L27}

\bibitem[\protect\citeauthoryear{{Hirano} et~al.,}{{Hirano}
  et~al.}{2020c}]{2020ApJ...899L..13H}
{Hirano} T.,  et~al., 2020c, \mn@doi [\apjl] {10.3847/2041-8213/aba6eb}, \href
  {https://ui.adsabs.harvard.edu/abs/2020ApJ...899L..13H} {899, L13}

\bibitem[\protect\citeauthoryear{{Howard} et~al.,}{{Howard}
  et~al.}{2023}]{HowardW2023}
{Howard} W.~S.,  et~al., 2023, \mn@doi [\apj] {10.3847/1538-4357/acfe75}, \href
  {https://ui.adsabs.harvard.edu/abs/2023ApJ...959...64H} {959, 64}

\bibitem[\protect\citeauthoryear{{Howell} et~al.,}{{Howell}
  et~al.}{2014}]{2014PASP..126..398H}
{Howell} S.~B.,  et~al., 2014, \mn@doi [\pasp] {10.1086/676406}, \href
  {http://adsabs.harvard.edu/abs/2014PASP..126..398H} {126, 398}

\bibitem[\protect\citeauthoryear{{Ida} \& {Lin}}{{Ida} \&
  {Lin}}{2008}]{2008ApJ...673..487I}
{Ida} S.,  {Lin} D.~N.~C.,  2008, \mn@doi [\apj] {10.1086/523754}, \href
  {https://ui.adsabs.harvard.edu/abs/2008ApJ...673..487I} {673, 487}

\bibitem[\protect\citeauthoryear{{Johnson}, {Cochran}, {Addison}, {Tinney}  \&
  {Wright}}{{Johnson} et~al.}{2017}]{2017AJ....154..137J}
{Johnson} M.~C.,  {Cochran} W.~D.,  {Addison} B.~C.,  {Tinney} C.~G.,
  {Wright} D.~J.,  2017, \mn@doi [\aj] {10.3847/1538-3881/aa8462}, \href
  {https://ui.adsabs.harvard.edu/abs/2017AJ....154..137J} {154, 137}

\bibitem[\protect\citeauthoryear{{Johnson} et~al.,}{{Johnson}
  et~al.}{2022}]{2022AJ....163..247J}
{Johnson} M.~C.,  et~al., 2022, \mn@doi [\aj] {10.3847/1538-3881/ac6271}, \href
  {https://ui.adsabs.harvard.edu/abs/2022AJ....163..247J} {163, 247}

\bibitem[\protect\citeauthoryear{{Kanodia} et~al.,}{{Kanodia}
  et~al.}{2022}]{Kanodia2022}
{Kanodia} S.,  et~al., 2022, \mn@doi [\apj] {10.3847/1538-4357/ac3e61}, \href
  {https://ui.adsabs.harvard.edu/abs/2022ApJ...925..155K} {925, 155}

\bibitem[\protect\citeauthoryear{{Klein} \& {Donati}}{{Klein} \&
  {Donati}}{2020}]{Klein2020}
{Klein} B.,  {Donati} J.~F.,  2020, \mn@doi [\mnras] {10.1093/mnrasl/slaa009},
  \href {https://ui.adsabs.harvard.edu/abs/2020MNRAS.493L..92K} {493, L92}

\bibitem[\protect\citeauthoryear{{Klein} et~al.,}{{Klein}
  et~al.}{2021}]{Klein2021}
{Klein} B.,  et~al., 2021, \mn@doi [\mnras] {10.1093/mnras/staa3702}, \href
  {https://ui.adsabs.harvard.edu/abs/2021MNRAS.502..188K} {502, 188}

\bibitem[\protect\citeauthoryear{{Kotani} et~al.,}{{Kotani}
  et~al.}{2018}]{2018SPIE10702E..11K}
{Kotani} T.,  et~al., 2018, in Ground-based and Airborne Instrumentation for
  Astronomy VII. p. 1070211, \mn@doi{10.1117/12.2311836}

\bibitem[\protect\citeauthoryear{{Kowalski}, {Allred}, {Carlsson}, {Kerr},
  {Tremblay}, {Namekata}, {Kuridze}  \& {Uitenbroek}}{{Kowalski}
  et~al.}{2022}]{Kowalski2022}
{Kowalski} A.~F.,  {Allred} J.~C.,  {Carlsson} M.,  {Kerr} G.~S.,  {Tremblay}
  P.-E.,  {Namekata} K.,  {Kuridze} D.,   {Uitenbroek} H.,  2022, \mn@doi
  [\apj] {10.3847/1538-4357/ac5174}, \href
  {https://ui.adsabs.harvard.edu/abs/2022ApJ...928..190K} {928, 190}

\bibitem[\protect\citeauthoryear{{Lin}, {Bodenheimer}  \& {Richardson}}{{Lin}
  et~al.}{1996}]{1996Natur.380..606L}
{Lin} D.~N.~C.,  {Bodenheimer} P.,   {Richardson} D.~C.,  1996, \mn@doi [\nat]
  {10.1038/380606a0}, \href {http://adsabs.harvard.edu/abs/1996Natur.380..606L}
  {380, 606}

\bibitem[\protect\citeauthoryear{{Lin}, {Ip}, {Hsiao}, {Chang}, {Song}  \&
  {Luo}}{{Lin} et~al.}{2023}]{LinC2023}
{Lin} C.-L.,  {Ip} W.-H.,  {Hsiao} Y.,  {Chang} T.-H.,  {Song} Y.-h.,   {Luo}
  A.~L.,  2023, \mn@doi [\aj] {10.3847/1538-3881/ace322}, \href
  {https://ui.adsabs.harvard.edu/abs/2023AJ....166...82L} {166, 82}

\bibitem[\protect\citeauthoryear{{Linsky}, {France}  \& {Ayres}}{{Linsky}
  et~al.}{2013}]{Linsky2013}
{Linsky} J.~L.,  {France} K.,   {Ayres} T.,  2013, \mn@doi [\apj]
  {10.1088/0004-637X/766/2/69}, \href
  {https://ui.adsabs.harvard.edu/abs/2013ApJ...766...69L} {766, 69}

\bibitem[\protect\citeauthoryear{{Linsky}, {Fontenla}  \& {France}}{{Linsky}
  et~al.}{2014}]{Linsky2014}
{Linsky} J.~L.,  {Fontenla} J.,   {France} K.,  2014, \mn@doi [\apj]
  {10.1088/0004-637X/780/1/61}, \href
  {https://ui.adsabs.harvard.edu/abs/2014ApJ...780...61L} {780, 61}

\bibitem[\protect\citeauthoryear{{Mann} et~al.,}{{Mann}
  et~al.}{2016}]{2016AJ....152...61M}
{Mann} A.~W.,  et~al., 2016, \mn@doi [\aj] {10.3847/0004-6256/152/3/61}, \href
  {https://ui.adsabs.harvard.edu/abs/2016AJ....152...61M} {152, 61}

\bibitem[\protect\citeauthoryear{{Martin}, {Lubow}, {Nixon}  \&
  {Armitage}}{{Martin} et~al.}{2016}]{2016MNRAS.458.4345M}
{Martin} R.~G.,  {Lubow} S.~H.,  {Nixon} C.,   {Armitage} P.~J.,  2016, \mn@doi
  [\mnras] {10.1093/mnras/stw605}, \href
  {https://ui.adsabs.harvard.edu/abs/2016MNRAS.458.4345M} {458, 4345}

\bibitem[\protect\citeauthoryear{{Martioli} et~al.,}{{Martioli}
  et~al.}{2020}]{2020A&A...641L...1M}
{Martioli} E.,  et~al., 2020, \mn@doi [\aap] {10.1051/0004-6361/202038695},
  \href {https://ui.adsabs.harvard.edu/abs/2020A&A...641L...1M} {641, L1}

\bibitem[\protect\citeauthoryear{{Masuda} \& {Winn}}{{Masuda} \&
  {Winn}}{2020}]{2020AJ....159...81M}
{Masuda} K.,  {Winn} J.~N.,  2020, \mn@doi [\aj] {10.3847/1538-3881/ab65be},
  \href {https://ui.adsabs.harvard.edu/abs/2020AJ....159...81M} {159, 81}

\bibitem[\protect\citeauthoryear{{Miyakawa}, {Hirano}, {Fukui}, {Mann},
  {Gaidos}  \& {Sato}}{{Miyakawa} et~al.}{2021}]{2021AJ....162..104M}
{Miyakawa} K.,  {Hirano} T.,  {Fukui} A.,  {Mann} A.~W.,  {Gaidos} E.,   {Sato}
  B.,  2021, \mn@doi [\aj] {10.3847/1538-3881/ac111d}, \href
  {https://ui.adsabs.harvard.edu/abs/2021AJ....162..104M} {162, 104}

\bibitem[\protect\citeauthoryear{{Montet} et~al.,}{{Montet}
  et~al.}{2020}]{2020AJ....159..112M}
{Montet} B.~T.,  et~al., 2020, \mn@doi [\aj] {10.3847/1538-3881/ab6d6d}, \href
  {https://ui.adsabs.harvard.edu/abs/2020AJ....159..112M} {159, 112}

\bibitem[\protect\citeauthoryear{{Nagasawa} \& {Ida}}{{Nagasawa} \&
  {Ida}}{2011}]{2011ApJ...742...72N}
{Nagasawa} M.,  {Ida} S.,  2011, \mn@doi [\apj] {10.1088/0004-637X/742/2/72},
  \href {http://adsabs.harvard.edu/abs/2011ApJ...742...72N} {742, 72}

\bibitem[\protect\citeauthoryear{{Newton} et~al.,}{{Newton}
  et~al.}{2019}]{2019ApJ...880L..17N}
{Newton} E.~R.,  et~al., 2019, \mn@doi [\apjl] {10.3847/2041-8213/ab2988},
  \href {https://ui.adsabs.harvard.edu/abs/2019ApJ...880L..17N} {880, L17}

\bibitem[\protect\citeauthoryear{{Ohno}, {Thao}, {Mann}  \& {Fortney}}{{Ohno}
  et~al.}{2022}]{2022ApJ...940L..30O}
{Ohno} K.,  {Thao} P.~C.,  {Mann} A.~W.,   {Fortney} J.~J.,  2022, \mn@doi
  [\apjl] {10.3847/2041-8213/ac9f3f}, \href
  {https://ui.adsabs.harvard.edu/abs/2022ApJ...940L..30O} {940, L30}

\bibitem[\protect\citeauthoryear{{Ohta}, {Taruya}  \& {Suto}}{{Ohta}
  et~al.}{2009}]{2009ApJ...690....1O}
{Ohta} Y.,  {Taruya} A.,   {Suto} Y.,  2009, \mn@doi [\apj]
  {10.1088/0004-637X/690/1/1}, \href
  {http://adsabs.harvard.edu/abs/2009ApJ...690....1O} {690, 1}

\bibitem[\protect\citeauthoryear{{Owen}}{{Owen}}{2019}]{Owen2019}
{Owen} J.~E.,  2019, \mn@doi [Annual Review of Earth and Planetary Sciences]
  {10.1146/annurev-earth-053018-060246}, \href
  {https://ui.adsabs.harvard.edu/abs/2019AREPS..47...67O} {47, 67}

\bibitem[\protect\citeauthoryear{{Owen} \& {Wu}}{{Owen} \&
  {Wu}}{2017}]{2017ApJ...847...29O}
{Owen} J.~E.,  {Wu} Y.,  2017, \mn@doi [\apj] {10.3847/1538-4357/aa890a}, \href
  {https://ui.adsabs.harvard.edu/abs/2017ApJ...847...29O} {847, 29}

\bibitem[\protect\citeauthoryear{{Palle} et~al.,}{{Palle}
  et~al.}{2020}]{2020A&A...643A..25P}
{Palle} E.,  et~al., 2020, \mn@doi [\aap] {10.1051/0004-6361/202038583}, \href
  {https://ui.adsabs.harvard.edu/abs/2020A&A...643A..25P} {643, A25}

\bibitem[\protect\citeauthoryear{{Pietrow} et~al.,}{{Pietrow}
  et~al.}{2023}]{Pietrow2023}
{Pietrow} A.~G.~M.,  et~al., 2023, \mn@doi [arXiv e-prints]
  {10.48550/arXiv.2309.03373}, \href
  {https://ui.adsabs.harvard.edu/abs/2023arXiv230903373P} {p. arXiv:2309.03373}

\bibitem[\protect\citeauthoryear{{Pizzolato}, {Maggio}, {Micela}, {Sciortino}
  \& {Ventura}}{{Pizzolato} et~al.}{2003}]{Pizzolato2003}
{Pizzolato} N.,  {Maggio} A.,  {Micela} G.,  {Sciortino} S.,   {Ventura} P.,
  2003, \mn@doi [\aap] {10.1051/0004-6361:20021560}, \href
  {https://ui.adsabs.harvard.edu/abs/2003A&A...397..147P} {397, 147}

\bibitem[\protect\citeauthoryear{{Plavchan} et~al.,}{{Plavchan}
  et~al.}{2020}]{2020Natur.582..497P}
{Plavchan} P.,  et~al., 2020, \mn@doi [\nat] {10.1038/s41586-020-2400-z}, \href
  {https://ui.adsabs.harvard.edu/abs/2020Natur.582..497P} {582, 497}

\bibitem[\protect\citeauthoryear{{Queloz}, {Eggenberger}, {Mayor}, {Perrier},
  {Beuzit}, {Naef}, {Sivan}  \& {Udry}}{{Queloz}
  et~al.}{2000}]{2000A&A...359L..13Q}
{Queloz} D.,  {Eggenberger} A.,  {Mayor} M.,  {Perrier} C.,  {Beuzit} J.~L.,
  {Naef} D.,  {Sivan} J.~P.,   {Udry} S.,  2000, \aap, \href
  {https://ui.adsabs.harvard.edu/abs/2000A&A...359L..13Q} {359, L13}

\bibitem[\protect\citeauthoryear{{Reiners}}{{Reiners}}{2009}]{Reiners2009}
{Reiners} A.,  2009, \mn@doi [\aap] {10.1051/0004-6361/200810257}, \href
  {https://ui.adsabs.harvard.edu/abs/2009A&A...498..853R} {498, 853}

\bibitem[\protect\citeauthoryear{{Schmidt}, {Kowalski}, {Hawley}, {Hilton},
  {Wisniewski}  \& {Tofflemire}}{{Schmidt} et~al.}{2012}]{Schmidt2012}
{Schmidt} S.~J.,  {Kowalski} A.~F.,  {Hawley} S.~L.,  {Hilton} E.~J.,
  {Wisniewski} J.~P.,   {Tofflemire} B.~M.,  2012, \mn@doi [\apj]
  {10.1088/0004-637X/745/1/14}, \href
  {https://ui.adsabs.harvard.edu/abs/2012ApJ...745...14S} {745, 14}

\bibitem[\protect\citeauthoryear{{Tamura} et~al.,}{{Tamura}
  et~al.}{2012}]{2012SPIE.8446E..1TT}
{Tamura} M.,  et~al., 2012, in Ground-based and Airborne Instrumentation for
  Astronomy IV. p. 84461T, \mn@doi{10.1117/12.925885}

\bibitem[\protect\citeauthoryear{{Thao} et~al.,}{{Thao}
  et~al.}{2023}]{2023AJ....165...23T}
{Thao} P.~C.,  et~al., 2023, \mn@doi [\aj] {10.3847/1538-3881/aca07a}, \href
  {https://ui.adsabs.harvard.edu/abs/2023AJ....165...23T} {165, 23}

\bibitem[\protect\citeauthoryear{{Tody}}{{Tody}}{1993}]{1993ASPC...52..173T}
{Tody} D.,  1993, in {Hanisch} R.~J.,  {Brissenden} R.~J.~V.,   {Barnes} J.,
  eds,  Astronomical Society of the Pacific Conference Series Vol. 52,
  Astronomical Data Analysis Software and Systems II. p.~173

\bibitem[\protect\citeauthoryear{{Vissapragada} et~al.,}{{Vissapragada}
  et~al.}{2021}]{Vissapragada2021}
{Vissapragada} S.,  et~al., 2021, \mn@doi [\aj] {10.3847/1538-3881/ac1bb0},
  \href {https://ui.adsabs.harvard.edu/abs/2021AJ....162..222V} {162, 222}

\bibitem[\protect\citeauthoryear{{Wilson} et~al.,}{{Wilson}
  et~al.}{2021}]{2021ApJ...911...18W}
{Wilson} D.~J.,  et~al., 2021, \mn@doi [\apj] {10.3847/1538-4357/abe771}, \href
  {https://ui.adsabs.harvard.edu/abs/2021ApJ...911...18W} {911, 18}

\bibitem[\protect\citeauthoryear{{Winn} \& {Fabrycky}}{{Winn} \&
  {Fabrycky}}{2015}]{Winn2015}
{Winn} J.~N.,  {Fabrycky} D.~C.,  2015, \mn@doi [\araa]
  {10.1146/annurev-astro-082214-122246}, \href
  {http://adsabs.harvard.edu/abs/2015ARA%26A..53..409W} {53, 409}

\bibitem[\protect\citeauthoryear{{Winn} et~al.,}{{Winn}
  et~al.}{2005}]{2005ApJ...631.1215W}
{Winn} J.~N.,  et~al., 2005, \mn@doi [\apj] {10.1086/432571}, \href
  {https://ui.adsabs.harvard.edu/abs/2005ApJ...631.1215W} {631, 1215}

\bibitem[\protect\citeauthoryear{{Winn}, {Fabrycky}, {Albrecht}  \&
  {Johnson}}{{Winn} et~al.}{2010}]{2010ApJ...718L.145W}
{Winn} J.~N.,  {Fabrycky} D.,  {Albrecht} S.,   {Johnson} J.~A.,  2010, \mn@doi
  [\apjl] {10.1088/2041-8205/718/2/L145}, \href
  {https://ui.adsabs.harvard.edu/abs/2010ApJ...718L.145W} {718, L145}

\bibitem[\protect\citeauthoryear{{Wirth} et~al.,}{{Wirth}
  et~al.}{2021}]{2021ApJ...917L..34W}
{Wirth} C.~P.,  et~al., 2021, \mn@doi [\apjl] {10.3847/2041-8213/ac13a9}, \href
  {https://ui.adsabs.harvard.edu/abs/2021ApJ...917L..34W} {917, L34}

\bibitem[\protect\citeauthoryear{{Wright}, {Newton}, {Williams}, {Drake}  \&
  {Yadav}}{{Wright} et~al.}{2018}]{Wright2018}
{Wright} N.~J.,  {Newton} E.~R.,  {Williams} P. K.~G.,  {Drake} J.~J.,
  {Yadav} R.~K.,  2018, \mn@doi [\mnras] {10.1093/mnras/sty1670}, \href
  {https://ui.adsabs.harvard.edu/abs/2018MNRAS.479.2351W} {479, 2351}

\bibitem[\protect\citeauthoryear{{Zhang}, {Knutson}, {Wang}, {Dai}  \&
  {Barrag{\'a}n}}{{Zhang} et~al.}{2022a}]{Zhang2022b}
{Zhang} M.,  {Knutson} H.~A.,  {Wang} L.,  {Dai} F.,   {Barrag{\'a}n} O.,
  2022a, \mn@doi [\aj] {10.3847/1538-3881/ac3fa7}, \href
  {https://ui.adsabs.harvard.edu/abs/2022AJ....163...67Z} {163, 67}

\bibitem[\protect\citeauthoryear{{Zhang} et~al.,}{{Zhang}
  et~al.}{2022b}]{Zhang2022a}
{Zhang} M.,  et~al., 2022b, \mn@doi [\aj] {10.3847/1538-3881/ac3f3b}, \href
  {https://ui.adsabs.harvard.edu/abs/2022AJ....163...68Z} {163, 68}

\bibitem[\protect\citeauthoryear{{Zhang}, {Knutson}, {Dai}, {Wang}, {Ricker},
  {Schwarz}, {Mann}  \& {Collins}}{{Zhang} et~al.}{2023}]{Zhang2023}
{Zhang} M.,  {Knutson} H.~A.,  {Dai} F.,  {Wang} L.,  {Ricker} G.~R.,
  {Schwarz} R.~P.,  {Mann} C.,   {Collins} K.,  2023, \mn@doi [\aj]
  {10.3847/1538-3881/aca75b}, \href
  {https://ui.adsabs.harvard.edu/abs/2023AJ....165...62Z} {165, 62}

\bibitem[\protect\citeauthoryear{{Zhou} et~al.,}{{Zhou}
  et~al.}{2020}]{2020ApJ...892L..21Z}
{Zhou} G.,  et~al., 2020, \mn@doi [\apjl] {10.3847/2041-8213/ab7d3c}, \href
  {https://ui.adsabs.harvard.edu/abs/2020ApJ...892L..21Z} {892, L21}

\makeatother
\end{thebibliography}








\bsp	
\label{lastpage}
\end{document}